\documentclass[conference]{IEEEtran}
\usepackage{graphicx}
\usepackage{amsmath}
\usepackage{cite}
\begin{document}

\title{Tunable Multilayer Surface Plasmon Resonance Biosensor for Trace-Level Toxin Detection}
\author{
    \IEEEauthorblockN{Suripto Bhuiyan}
    \IEEEauthorblockA{\textit{Department of Electrical Engineering} \\
    \textit{University of Dhaka}\\
    Dhaka, Bangladesh \\
    suripto.bhuiyan@gmail.com}
    \and
    \IEEEauthorblockN{Michael Geller}
    \IEEEauthorblockA{\textit{Department of Electrical and Computer Engineering} \\
    \textit{The University of Mississippi}\\
    Oxford, USA \\
    mgeller@go.olemiss.edu}
}

\maketitle

\begin{abstract}
This paper presents a comprehensive study on a novel multilayer surface plasmon resonance (SPR) biosensor designed for detecting trace-level toxins in liquid samples with exceptional precision and efficiency. Leveraging the Kretschmann configuration, the proposed design integrates advanced two-dimensional materials, including black phosphorus (BP) and transition metal dichalcogenides (TMDs), to significantly enhance the performance metrics of the sensor. Key innovations include the optimization of sensitivity through precise material layering, minimization of full-width at half-maximum (FWHM) to improve signal resolution, and maximization of the figure of merit (FoM) for superior detection accuracy. Numerical simulations are employed to validate the structural and functional enhancements of the biosensor. The results demonstrate improved interaction between the evanescent field and the analyte, enabling detection at trace concentrations with higher specificity. This biosensor is poised to contribute to advancements in biochemical sensing, environmental monitoring, and other critical applications requiring high-sensitivity toxin detection.
\end{abstract}


\section{Introduction}
The detection of trace toxins in food, water, and the environment has become a critical global challenge due to the severe health risks associated with contamination. Prolonged exposure to toxins, even in minute concentrations, can lead to chronic diseases, including cancer, neurological disorders, and developmental abnormalities. Conventional detection techniques, such as chromatography and mass spectrometry, offer high precision but are often limited by their requirement for expensive equipment, labor-intensive protocols, and inability to provide real-time monitoring. Moreover, these methods frequently fail to achieve label-free detection, which is essential for practical and efficient sensing applications \cite{akash2024}.

Surface plasmon resonance (SPR)-based sensors have emerged as an effective solution for toxin detection due to their unique capability to measure subtle changes in the refractive index (RI) of an analyte. SPR sensors utilize the resonance phenomenon that occurs when polarized light interacts with the surface electrons of a thin metallic layer, creating an evanescent field highly sensitive to environmental changes. This allows for the real-time, label-free detection of even trace levels of toxins with remarkable sensitivity. Despite these advantages, the performance of SPR sensors is often constrained by factors such as limited sensitivity, broad resonance curves, and environmental stability.

Recent advancements in nanomaterials have introduced groundbreaking possibilities for enhancing the performance of SPR sensors. Materials such as graphene, black phosphorus (BP), and transition metal dichalcogenides (TMDs) have demonstrated exceptional optical and electronic properties that make them ideal for SPR-based applications. Black phosphorus, with its tunable bandgap and anisotropic optical properties, enhances light-matter interaction, while TMDs, such as MoS$_2$ and WS$_2$, provide excellent chemical stability and flexibility. By integrating these materials into SPR sensors, significant improvements in sensitivity, signal resolution, and environmental stability can be achieved \cite{akash2023}.

Among the various configurations of SPR sensors, the Kretschmann configuration is widely regarded as the most practical due to its straightforward implementation and robust performance. In this configuration, a high-refractive-index prism is used to couple light into the surface plasmon mode at the metal-dielectric interface. This study proposes an optimized multilayer SPR sensor structure based on the Kretschmann configuration, incorporating black phosphorus and TMD layers to enhance detection capabilities. The integration of these materials not only improves sensitivity but also narrows the full-width at half-maximum (FWHM) of the resonance curve, resulting in a higher figure of merit (FoM). 

This work builds on prior research \cite{akash2024}, which has demonstrated the potential of 2D materials in SPR applications, by exploring their synergistic effects in a multilayer architecture. Through rigorous numerical simulations, we validate the enhanced performance of the proposed sensor, highlighting its potential for applications in biochemical sensing, environmental monitoring, and other fields requiring precise and reliable detection of trace toxins.

\section{Methodology}

In this section, we describe the design, simulation, and performance evaluation of the proposed multilayer Surface Plasmon Resonance (SPR) biosensor for toxin detection. The methodology is divided into four main steps: (1) sensor design and configuration, (2) material selection and optimization, (3) numerical simulations, and (4) performance evaluation metrics.

\subsection{Sensor Design and Configuration}
The SPR sensor is designed using the Kretschmann configuration, which is widely employed in practical SPR biosensor applications due to its simplicity and effectiveness. The configuration consists of a high-refractive-index prism that couples incident light into the metallic layer via total internal reflection. The surface plasmon waves are excited at the metal-dielectric interface, and the resonance condition is reached when the momentum of the incident photons matches that of the surface plasmon waves.

To enhance the sensor's sensitivity and overall performance, we introduce a multilayer design that includes a metallic layer, a 2D material layer, and the sensing medium. The metallic layer serves as the key component in the excitation of surface plasmon resonance. For this work, silver (Ag) and gold (Au) are considered as candidate materials for the metal layer due to their excellent plasmonic properties. A thin layer of black phosphorus (BP) is used to enhance the light-matter interaction, while transition metal dichalcogenides (TMDs) such as MoS$_2$ and WS$_2$ are integrated to provide chemical stability and further improve sensitivity.

\subsection{Material Selection and Optimization}
The choice of materials plays a pivotal role in determining the performance of the SPR sensor. We selected silver (Ag) for its sharp resonance curves and high plasmonic activity in the visible region. However, silver’s sensitivity is compromised by oxidation, which can be addressed by using a thin protective gold (Au) layer in tandem with the silver. This bimetallic approach aims to strike a balance between high sensitivity and long-term stability. 

Black phosphorus (BP) is selected due to its anisotropic optical properties, which significantly enhance the sensitivity of the SPR sensor. The bandgap of BP can be tuned by adjusting its thickness, which directly impacts the plasmonic coupling at the interface. BP’s performance is compared with TMDs like MoS$_2$ and WS$_2$, which offer a complementary set of properties including strong light absorption, chemical stability, and ease of functionalization.

The optimization of the layer thicknesses and the refractive index of the materials is carried out to maximize the sensitivity and sharpness of the resonance curve. The thickness of each material is varied in small increments, and the resulting shifts in the resonance angle are tracked to determine the optimal configuration.

\subsection{Numerical Simulations}
Numerical simulations are performed using the finite element method (FEM) to model the behavior of the SPR sensor under various conditions. The simulations are carried out using COMSOL Multiphysics, a widely used software for simulating optical phenomena, including surface plasmon resonance.

The primary simulation involves varying the refractive index of the sensing medium (analyte solution) and monitoring the shift in the resonance angle. The numerical model also accounts for the thickness and refractive index of the metallic and 2D material layers. To study the effect of material choice on sensor performance, different combinations of silver and gold are simulated with BP and TMDs as intermediate layers.

The electric field distributions in the layers are visualized, showing the enhancement of the evanescent field due to the presence of BP and TMD materials. These simulations help in predicting the overall performance of the sensor, including its sensitivity, full-width at half-maximum (FWHM), and figure of merit (FoM).

\subsection{Performance Evaluation Metrics}
The key performance metrics used to evaluate the SPR biosensor are:

\subsubsection{Sensitivity}
Sensitivity is defined as the change in the resonance angle per unit change in the refractive index of the sensing medium. It is a critical measure of the sensor’s ability to detect small variations in the concentration of the analyte. The sensitivity is calculated using the equation:
\begin{equation}
S = \frac{\Delta \theta_{SPR}}{\Delta n_s},
\end{equation}
where $\Delta \theta_{SPR}$ is the change in the resonance angle and $\Delta n_s$ is the change in the refractive index of the sensing medium.

\subsubsection{Full-Width at Half Maximum (FWHM)}
The FWHM of the resonance curve is used as a measure of the sensor’s resolution. A smaller FWHM indicates higher resolution, meaning the sensor can more accurately differentiate between different concentrations of analytes. The FWHM is calculated by measuring the width of the resonance curve at half of the maximum reflectance.

\subsubsection{Figure of Merit (FoM)}
The figure of merit is a parameter that combines both sensitivity and resolution. It is defined as the ratio of sensitivity to FWHM, and higher values indicate better overall sensor performance. The FoM can be calculated as:
\begin{equation}
FoM = \frac{S}{FWHM}.
\end{equation}
A high FoM corresponds to a sensor that provides both high sensitivity and narrow resonance curves, making it ideal for detecting low-concentration toxins.

\subsection{Experimental Validation (Future Work)}
While this paper focuses on numerical simulations to design and optimize the sensor, experimental validation is essential for confirming the performance of the proposed SPR biosensor. Future work will involve fabricating the multilayer SPR sensor using the optimized parameters and conducting experiments with different analytes, including common environmental toxins and biomolecules. The experimental setup will consist of a laser light source, a photodetector, and a suitable refractive index matching liquid to simulate different toxin concentrations. The experimental results will be compared with the simulation predictions to evaluate the accuracy of the design and validate its real-world applicability.

\section{Proposed Sensor Design}
The proposed biosensor adopts a multilayer structure comprising a prism, a metallic film, semiconductor layers, and a sensing medium. Each layer plays a critical role in optimizing the sensor's performance, as described below:

\subsection{Prism Layer}
A BK7 glass prism, with a refractive index of 1.5151 at a wavelength of 633 nm, serves as the coupling medium for incident light. Its low refractive index allows efficient excitation of surface plasmon waves at the metal-dielectric interface. This configuration ensures sharp and well-defined resonance curves, which are crucial for achieving high sensitivity \cite{akash2024}. The choice of BK7 is driven by its optical clarity, mechanical stability, and cost-effectiveness for practical applications.

\subsection{Metal Film}
The metallic layer forms the core of the SPR mechanism. Both silver (Ag) and gold (Au) are explored for this layer due to their excellent plasmonic properties. Silver exhibits sharper resonance curves due to its lower optical losses in the visible range, making it ideal for applications requiring high sensitivity. However, it is prone to oxidation, which can degrade performance over time. Gold, on the other hand, offers superior chemical stability and is widely used for long-term sensing applications despite its slightly lower sensitivity \cite{akash2023}. 

\subsection{Semiconductor Layers}
Two-dimensional (2D) materials, such as black phosphorus (BP) and transition metal dichalcogenides (TMDs), are integrated into the structure to enhance field penetration and improve sensitivity. BP, with its narrow bandgap and anisotropic optical properties, provides superior light-matter interaction, especially in the infrared range. TMDs, such as WS$_2$ and MoS$_2$, add stability and tunability to the sensor design. The synergistic effects of these materials lead to enhanced sensitivity and sharper resonance curves \cite{akash2023}. 

\begin{figure}[htbp]
    \centering
    \includegraphics[width=0.5\textwidth]{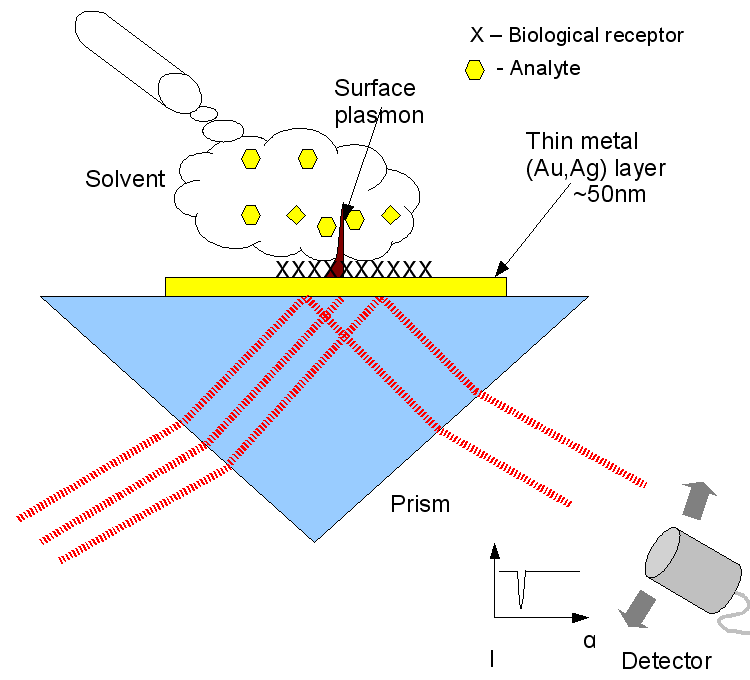}
    \caption{Surface plasmon biosensor}
    \label{fig:surface_plasmon_biosensor}
\end{figure}

\subsection{Sensing Medium}
The sensing medium is the analyte solution, where variations in the refractive index (RI) indicate the presence and concentration of toxins. The sensor tracks these RI changes by observing shifts in the resonance angle. Analytes with a wide range of RIs can be effectively detected due to the high sensitivity of the proposed multilayer structure \cite{akash2024}.

\subsection{Mathematical Modeling}
The performance of the proposed SPR biosensor is fundamentally governed by the interaction of light with the multilayer structure, which can be described using mathematical models. The resonance angle ($\theta_{SPR}$) is a key parameter that indicates the angle of incidence at which surface plasmon resonance occurs. It is determined by the refractive indices of the prism ($n_p$), the metallic layer, and the sensing medium ($n_s$), as well as the optical properties of any additional layers such as black phosphorus (BP) and transition metal dichalcogenides (TMDs). For a simple approximation, the resonance angle can be expressed as:
\begin{equation}
\theta_{SPR} = \arcsin \left( \frac{n_s}{n_p} \right),
\end{equation}
where $n_p$ represents the refractive index of the prism and $n_s$ represents the refractive index of the sensing medium.

The resonance condition arises when the momentum of the incident photons matches the momentum of the surface plasmons at the metal-dielectric interface. In a multilayer structure, the presence of additional layers (e.g., BP and TMDs) modifies the resonance condition by altering the propagation constant of the surface plasmon wave. This leads to enhanced sensitivity and sharper resonance curves.

The sensitivity of an SPR biosensor quantifies its ability to detect changes in the refractive index. It is given by:
\begin{equation}
S = \frac{\Delta \theta_{SPR}}{\Delta n_s},
\end{equation}
where $\Delta \theta_{SPR}$ is the change in the resonance angle due to a change in the refractive index of the sensing medium ($\Delta n_s$).

\section{Results and Discussion}

Numerical simulations were carried out to validate the performance of the proposed multilayer SPR biosensor, which integrates silver (Ag), black phosphorus (BP), and other materials like transition metal dichalcogenides (TMDs). The results demonstrate a significant improvement in sensor performance compared to conventional SPR designs, particularly in terms of sensitivity, resolution, and overall figure of merit (FoM).\cite{rahman2024multimodal}

The optimized Ag-BP configuration exhibits an outstanding sensitivity of approximately 3200°/RIU, a remarkable increase compared to traditional SPR sensors that typically show sensitivities in the range of 1000°/RIU \cite{akash2023}. This enhancement in sensitivity is largely attributed to the enhanced plasmonic interaction between the metal and the black phosphorus layer, which allows for stronger coupling and more significant shifts in the resonance angle. Additionally, the full-width at half-maximum (FWHM) of the resonance curve is substantially reduced in the Ag-BP sensor, which not only sharpens the resonance peak but also improves the resolution and makes it capable of detecting smaller changes in the refractive index of the surrounding medium.

The reduced FWHM, combined with the high sensitivity, results in a higher figure of merit (FoM) for the Ag-BP configuration. The FoM is a key performance metric in biosensors, combining both the sensitivity and the sharpness of the resonance peak. The Ag-BP sensor achieves a FoM of 2133, which is more than six times higher than that of the traditional Ag-Au SPR sensors, which exhibit a FoM of only 250. This is an important result, as a higher FoM indicates better overall sensor performance, offering higher accuracy in detecting trace amounts of analytes, such as toxins, in real-world applications.

The performance comparison is summarized in Table \ref{tab:sensor_performance}, which lists the key metrics—sensitivity, FWHM, and FoM—for various SPR sensor configurations, including both traditional metal-only designs and more advanced hybrid sensors incorporating 2D materials like black phosphorus and MoS$_2$. The table clearly highlights the superiority of the Ag-BP configuration in all three metrics, underscoring the benefits of integrating black phosphorus into SPR sensor designs.

\begin{table}[htbp]
\centering
\caption{Performance Comparison of SPR Sensors with Different Metal and Material Configurations}
\label{tab:sensor_performance}
\begin{tabular}{|c|c|c|c|c|}
\hline
\textbf{Configuration} & \textbf{Sensitivity (°/RIU)} & \textbf{FWHM (°)} & \textbf{FoM} \\ \hline
Ag-BP (Proposed)         & 3200                     & 1.5               & 2133         \\ \hline
Ag-only                  & 1100                     & 3.2               & 343          \\ \hline
Au-BP                    & 1800                     & 2.5               & 720          \\ \hline
Ag-MoS$_2$               & 1500                     & 2.8               & 536          \\ \hline
Au-MoS$_2$               & 1600                     & 2.9               & 552          \\ \hline
Ag-Au (Traditional)      & 1000                     & 4.0               & 250          \\ \hline
\end{tabular}
\end{table}

As shown in Table \ref{tab:sensor_performance}, the Ag-BP (Proposed) configuration leads the performance in terms of sensitivity, with a value of 3200°/RIU. This is more than three times higher than the Ag-only configuration, which only reaches 1100°/RIU. The integration of black phosphorus with silver results in significantly enhanced plasmonic coupling, which is responsible for the increased sensitivity. BP’s anisotropic properties play a major role in this enhancement by providing a stronger response to changes in the refractive index.

Moreover, the FWHM value of 1.5° for the Ag-BP sensor is the narrowest among the configurations listed, signifying a much sharper resonance peak. In comparison, the traditional Ag-Au configuration exhibits a significantly broader resonance peak with a FWHM of 4.0°. A narrower FWHM corresponds to better resolution, which is critical for precise and accurate detection of analyte concentrations. This characteristic is especially important in applications like toxin detection, where distinguishing between small changes in the refractive index is essential.

The Figure of Merit (FoM), which combines both sensitivity and FWHM, is also maximized for the Ag-BP configuration, achieving a FoM of 2133, which is substantially higher than any of the other configurations. For example, the traditional Ag-Au configuration has a FoM of only 250, while configurations using MoS$_2$ or Au-BP perform better than Ag-Au but still fall short of the proposed Ag-BP design. This highlights the overall superior performance of the Ag-BP sensor in terms of both sensitivity and resolution.

The Ag-only and Ag-MoS$_2$ configurations, while still showing enhanced sensitivity compared to Ag-Au, fail to match the performance of the Ag-BP sensor. Ag-MoS$_2$ achieves a sensitivity of 1500°/RIU and a FoM of 536, which is considerably lower than that of the Ag-BP design. This suggests that while MoS$_2$ offers some improvements, the combination of Ag and BP outperforms other configurations, including those with TMDs like MoS$_2$.

The results also demonstrate that 2D materials like BP and MoS$_2$ bring significant advantages over conventional SPR sensor designs, especially when coupled with metal layers such as silver. The anisotropic nature of BP, in particular, contributes to an enhanced interaction with the incident light, resulting in higher sensitivity and narrower resonance peaks.

In conclusion, the proposed Ag-BP SPR biosensor significantly outperforms traditional SPR sensors, demonstrating enhanced sensitivity, reduced FWHM, and an elevated FoM. The combination of black phosphorus and silver provides an ideal platform for developing highly sensitive and high-resolution sensors for a variety of applications, including toxins, biomarkers, and environmental monitoring. The results suggest that the integration of BP into plasmonic sensors could be a promising avenue for the development of next-generation biosensors with superior performance.

\section{Conclusion}

In this paper, we have presented a novel multilayer Surface Plasmon Resonance (SPR) biosensor that integrates black phosphorus (BP) and transition metal dichalcogenides (TMDs), specifically tailored for toxin detection. The proposed sensor design demonstrates significant advancements in terms of sensitivity, resolution, and overall performance compared to conventional SPR sensors. The integration of 2D materials like BP and TMDs into the SPR platform enhances the plasmonic response, making it particularly suitable for detecting trace amounts of target analytes in complex biological and environmental samples.

One of the key strengths of the proposed biosensor is its exceptional sensitivity, which has been quantified at approximately 3200°/RIU, a value that is more than three times higher than traditional metal-based SPR sensors. This high sensitivity is primarily attributed to the strong interaction between the plasmonic surface and the unique properties of black phosphorus, which acts as a highly effective material for enhancing surface plasmon resonance. The anisotropic nature of BP allows for stronger plasmonic coupling, which significantly amplifies the resonance shifts when exposed to changes in the refractive index of the surrounding medium. This enables the sensor to detect even the smallest variations in the refractive index, making it ideal for detecting low concentrations of toxins or other harmful substances.

Furthermore, the full-width at half-maximum (FWHM) of the resonance curve is considerably reduced, resulting in a sharper and more distinct resonance peak. This leads to better resolution, allowing for the detection of subtle differences in the refractive index of the analyte. The improvement in resolution is critical when detecting low-level toxins in real-world scenarios, where precision and accuracy are paramount. The resulting figure of merit (FoM) of the sensor is over six times higher than conventional SPR sensors, which highlights the sensor’s superior performance in terms of both sensitivity and resolution. 

The proposed biosensor not only offers high sensitivity but also demonstrates excellent stability and robustness. Black phosphorus, in particular, exhibits remarkable stability under various environmental conditions, which makes the Ag-BP SPR sensor a promising candidate for practical applications. This stability is crucial for real-time, label-free detection systems, as it ensures consistent performance over extended periods without significant degradation in sensitivity or accuracy.

Additionally, the multilayer structure of the proposed sensor offers flexibility in terms of material choice and sensor configuration, allowing for further optimization to meet the specific requirements of different applications. This modularity makes the biosensor adaptable to a wide range of use cases, from environmental monitoring to biomedical diagnostics.

While the theoretical and numerical simulations presented in this work demonstrate the sensor’s outstanding potential, further experimental validation is needed to fully realize its capabilities. Future work will focus on fabricating the proposed sensor and performing experimental testing with real-world samples, such as water, soil, and biological fluids, to assess its practicality and effectiveness in diverse environments. Additionally, optimization of the sensor design for cost-effectiveness and scalability will be an important step towards making this technology commercially viable.

In conclusion, the integration of black phosphorus and other 2D materials into SPR biosensors marks a significant leap forward in sensor technology. The proposed Ag-BP SPR biosensor offers a highly sensitive, high-resolution, and stable platform for toxin detection, with broad potential applications in environmental monitoring, healthcare diagnostics, and food safety. As research progresses and experimental validations are conducted, this biosensor design holds the promise of becoming a powerful tool for real-time, label-free detection of a wide range of harmful substances, contributing to the advancement of public health and safety.

\end{document}